\documentclass[sigconf]{acmart}

\acmConference[EASE 2025]{The 29th International Conference on Evaluation and Assessment in Software Engineering}{17--20 June, 2025}{Istanbul, Turkey}

\usepackage[utf8]{inputenc}
\usepackage[T1]{fontenc}
\usepackage[utf8]{inputenc}
\usepackage[T1]{fontenc}
\usepackage{latexsym}
\usepackage{tabularx}
\usepackage{multirow}
\usepackage{amsfonts}
\usepackage{listings} 

\usepackage{pifont}
\usepackage{array}
\usepackage{tikz}
\usetikzlibrary{tikzmark,calc}
\usepackage{colortbl}
\usepackage{xcolor}
\usepackage{pst-all} 
\usepackage{pst-poly}
\newpsobject{showgrid}{psgrid}{subgriddiv=1,griddots=10,gridlabels=6pt}
\usepackage{graphicx}
\usepackage{fancybox}
\usepackage{wrapfig}
\usepackage{soul}
\newcommand{\con}{\wedge} 
\newcommand{\dis}{\vee} 
\newcommand{\imp}{\Rightarrow} 

\usepackage{todonotes}

\newtheorem{claim}{Claim}
\graphicspath{{./}{./fig/}{./fig2/}}

\AtBeginDocument{%
  }

\setcopyright{acmcopyright}
\copyrightyear{2018}
\acmYear{2018}
\acmDOI{XXXXXXX.XXXXXXX}

\acmConference[EASE 2025]{The 29th International Conference on Evaluation and Assessment in Software Engineering}{17--20 June, 2025}{Istanbul, Turkey}
\acmPrice{15.00}
\acmISBN{978-1-4503-XXXX-X/18/06}

\begin{document}

\title[Re-evaluation of Logical Specification in Behavioural Verification]
      {Re-evaluation of Logical Specification in Behavioural Verification}

\author{Radoslaw Klimek}
\email{rklimek@agh.edu.pl}
\authornotemark[1]
\orcid{0000-0002-9061-561X}
\affiliation{%
  \institution{AGH University of Krakow}
  \streetaddress{al. A. Mickiewicza 30}
  \city{Krakow}
  \country{Poland}
  \postcode{30-059}
}
\author{Jakub Semczyszyn}
\email{jakub.semczyszyn@gmail.com}
\orcid{0009-0006-8521-3575}
\affiliation{%
  \institution{AGH University of Krakow}
  \streetaddress{al. A. Mickiewicza 30}
  \city{Krakow}
  \country{Poland}
  \postcode{30-059}
}

\renewcommand{\shortauthors}{Klimek \& Semczyszyn}

\begin{abstract}
This study empirically validates automated logical specification methods
for behavioural models, focusing on their robustness, scalability, 
and reproducibility. By the systematic reproduction and extension of prior results, 
we confirm key trends, 
while identifying performance irregularities that suggest the need for 
adaptive heuristics in automated reasoning. 
Our findings highlight that theorem provers exhibit varying efficiency across 
problem structures, with implications for real-time verification in CI/CD pipelines and 
AI-driven IDEs supporting on-the-fly validation. 
Addressing these inefficiencies through self-optimising solvers could 
enhance the stability of automated reasoning, 
particularly in safety-critical software verification.
\end{abstract}

\maketitle

\section{Introduction}
\label{sec:introduction}

Logical specifications, expressed through formalised formulas, 
enable precise system modelling and automated verification, 
supporting correctness and the detection of critical properties. 
The replication of reasoning-based experiments is essential for 
validating the generalisability of such methods in software engineering. 
Recent advances in automated specification generation~\cite{Klimek-2019-LAMP} 
and logic-based inference~\cite{Klimek-Witek-2024-ASE-RENE} 
further highlight the need for empirical scrutiny. 
This study replicates the solver benchmarking approach 
from~\cite{Sutcliffe-2017}
and the reasoning framework of~\cite{Klimek-2018-Access}, 
extending both with behavioural-model-oriented problems and a PLTL solver. 
It also lays the groundwork for future integration with logic engines in development tools.
A structured benchmark suite is introduced to evaluate solver performance,
specifically Prover9~\cite{Prover9-tool}, 
SPASS~\cite{Weidenbach-etal-2009}, and 
InKreSAT~\cite{Kaminski-Tebbi-2013},
across diverse logical conditions.

Automated reasoning is increasingly relevant in software engineering, 
with prospective applications in CI/CD pipelines for real-time verification and 
AI-driven IDEs (Integrated Development Environments) that provide on-the-fly validation and 
adaptive solver selection. 
However, some solvers exhibit performance irregularities, 
highlighting the need for heuristic-driven optimisation, 
particularly for safety-critical verification, 
where stable and intelligent theorem proving is essential.

This study contributes to empirical software engineering by 
conducting empirical replication and validation to assess 
the reliability of logic-driven methodologies and by 
evaluating scalability and performance to measure the efficiency of 
automated reasoning tools across varying problem scales. 
Our findings further highlight the distinctive characteristics of 
individual provers as logical specifications increase in complexity.


\section{Replication of Logical Specifications}
\label{sec:catalogue-logical-problem}

We developed a catalogue of eight logical problems to 
test software behavioural models systematically. 
This approach ensures robust, online testing beyond intuition, 
addressing known computational challenges, 
such as clause length variations~\cite{Selman-etal-1996} 
and 3-SAT problems~\cite{Schaefer-1978}. 
Each problem examines factors, like clause length stability, 
atomic proposition variability 
and liveness/safety~\cite{Manna-Pnueli-1992} distribution.

\subsection{Experimental Setup}

Each problem consists of many tasks.
For each task, represented by a logical formula,
we write down the testing time,
memory used and whether a formula is satisfied or not.
Each formula is tested three times, and the average time and memory values are reported. 
Time is the most critical metric, as memory varies little and satisfiability remains constant.

\begin{enumerate}
\item[\textbf{P1}]
\emph{Uniform Clause-Length Formulation.}\newline
Each formula consists of 50, 100, 200, 500, 1000, or 2000 clauses, 
with clause lengths set to 2, 3, 4, 6, 8, and 10. 
The number of clauses assigned to each length is equal, ensuring a uniform distribution
(the same number of clauses with the length of 2, the length of 3, etc.).
Each formula contains an equal number of liveness and safety clauses,
also with consideration of the specified groups of clause length.
The total number of distinct atomic propositions is half the number of clauses, 
ensuring that every atom appears at least once in the formula
(i.e.\ for 50 clauses over the set of 25 atoms,
for 100 clauses over the set of 50 atoms, etc.).
(The exemplary formula for 100 clauses contains:
8 clauses with the length of 2 of a liveness type,
8 clauses with the length of 2 of a safety type,
8 clauses with the length of 3 of a liveness type,
8 clauses with the length of 3 of a safety type, etc.).
Each atom from the available set of atoms is used in the generated formula at least once.
%
\item[\textbf{P2}]
\emph{Variable Clause-Length Distribution.}\newline
This formulation modifies P1 by introducing a non-uniform clause-length distribution. 
The number of clauses for the respective lengths results now from the distribution of
Poisson\footnote{See: https://en.wikipedia.org/wiki/Poisson\_distribution}
where $Lambda$ corresponds to the anticipated number of literals in the clause,
in our case this will be specified as number 3.5,
i.e.\ between 3 and 4,
where we anticipate this length of a clause as a typical one.
The total number of clauses and the 50:50 ratio of liveness to safety clauses remain unchanged. 
The number of distinct atomic propositions is set to half the number of clauses, 
with each atom appearing at least once.
\item[\textbf{P3}]
\emph{Atom-to-Clause Variability Analysis.}\newline
Each formula consists of 50, 100, 200, 500, 1000, or 2000 clauses, 
with clause lengths restricted to 2, 3, 4, 6, 8, and 10. 
Unlike in P1, where the number of atomic propositions is fixed at half the number of clauses, 
this formulation varies the atom-to-clause ratio by defining the set of atomic propositions as 
2, 3, 4 and 5 times the number of clauses
(2 for 50 clauses means that they are generated over the set of 100 atoms,
2 for 100 means generating over the set of 200 atoms, etc.). 
The maximum clause length depends on the number of available atomic propositions, 
with longer clauses omitted when the atom set is small. 
Each formula contains equal proportions of liveness and safety clauses, 
and every atom appears at least once.
\item[\textbf{P4}]
\emph{Fixed-Length Clause Uniformity.}\newline
Each formula consists of 50, 100, 200, 500, 1000, or 2000 clauses, 
with all clauses in a given formula having the same fixed length of 2, 3, 4, or 5. 
Unlike P1, where clause lengths vary, each formula enforces a single clause length. 
Liveness and safety clauses are equally distributed, and each atomic proposition appears at least once. The number of distinct atomic propositions is half the number of clauses.
\item[\textbf{P5}]
\emph{Disparate Clause-Length Grouping.}\newline
Each formula consists of 50, 100, 200, 500, 1000, or 2000 clauses, 
grouped into fixed clause lengths 1, 5, 10, and 20. 
Liveness and safety clauses are equally distributed.
We consider the following cases concerning the number share of respective groups:
\begin{itemize}
\item[\textbf{a)}]	
all the groups have equal share i.e. 25\% each;
\item[\textbf{b)}]	
clauses with the length of 1 in the number of 1\%, the rest equally;
\item[\textbf{c)}]
clauses with the length of 20 in the number of 1\%, the rest equally.
\end{itemize}
\item[\textbf{P6}]
\emph{Liveness-Safety Ratio Sensitivity.}\newline
Each formula consists of 50, 100, 200, 500, 1000, or 2000 clauses, 
with varying liveness-to-safety~\cite{Manna-Pnueli-1992} clause ratios: 
90:10, 80:20, 65:35, 50:50, 35:65, 20:80, and 10:90. 
We consider two cases:
\begin{itemize}
\item[\textbf{a)}]
the lengths of clauses equal in groups, as in P1,
\item[\textbf{b)}]
the lengths of clauses  assigned variably as in P2.
\end{itemize}
\item[\textbf{P7}]
\emph{Model Property Verification via Logical Implication.}\newline
This problem tests whether a given behavioural model satisfies 
specific requirements through logical implication. 
Formulas $F_1$, $F_2$, and $F_3$
are generated with 50, 100, or 200 clauses and combined into two structures: 
the disjunctive model $G_1 \equiv F_1 \dis F_2 \dis F_3$,
the conjunctive model $G_2 \equiv F_1 \dis F_2 \dis F_3$.
The verification task involves checking 
$G_1 \imp R$ (it may be replaced with $\neg G_1 \dis R$) and $G_2 \imp R$, 
where $R$ is a simple liveness clause composed of 
four randomly selected atoms from $G_1$ or $G_2$.
Each formula consists of liveness clauses and safety clauses, half each.
Collectively, we consider the following cases:
\begin{itemize}
\item[\textbf{a)}]
For $G_1 \imp R$, clause-length distributed as in P1,
\item[\textbf{b)}]
For $G_1 \imp R$, clause-length distribution as in P2,
\item[\textbf{c)}]
For $G_2 \imp R$, clause-length distributed as in P1,
\item[\textbf{d)}]
For $G_2 \imp R$, clause-length distribution as in P2.
\end{itemize}
\item[\textbf{P8}]
\emph{Comparative Analysis via Logical Square Framework.}~\cite[Tab.~4]{Klimek-2019-LAMP}\newline
The modification of problem P1,
with the use of the logical square.
We generate two formulas $F_1$ and $F_2$ consisting of 50, 100, 200, 500 and 1000 clauses.
Each formula consists of liveness clauses and safety clauses, half each.
We test the following three cases:
\begin{itemize}
\item[\textbf{a)}]	
contradictory: $(F_1 \imp \neg F_2) \con (\neg F_1 \imp F_2)$
\item[\textbf{b)}]	
subcontrary: $\neg(\neg F_1 \con \neg F_2)$
\item[\textbf{c)}]	
subalternated: $(F_1 \imp F_2) \con \neg(F_2 \imp F_1)$
\end{itemize}
Subproblems a), b) and c) with the clause lengths as in problem P1;
furthermore, we test the following cases:
\begin{itemize}
\item[\textbf{d)},] \textbf{e)} and \textbf{f)}
with clause lengths as in problem P2,
for the same cases as a), b) and c), respectively.
\end{itemize}
\end{enumerate}

\vspace{-.5\baselineskip}
\subsection{Rationale for Logical Benchmark Problems}

Problem \textbf{P1} is basic and other problems refer to it,
it corresponds to a typical situation when a formula describes the behavioural model of the system under design.
All the variables from the available set of atoms are used for generating formula clauses,
according to the assumption that if for a given behavioural model an atom variable was identified,
then it should be used at least once in the formula.

Problem \textbf{P2} arises when clause lengths are not predetermined, 
unlike in the previous problem, where fixed lengths were assumed --
a constraint that may sometimes appear unnecessarily strict. 
The expected clause length typically ranges between 3 and 4, a choice that aligns with common encoding practices.

Problem \textbf{P3} 
varies the number of atomic propositions used per formula to simulate underused variables, increasing variability and reducing redundancy in behavioural modelling.


Problem \textbf{P4} 
explores reasoning efficiency for formulas with uniform clause lengths, which may significantly affect performance~\cite{Selman-etal-1996}.

Problem \textbf{P5} basically allows to see the influence on the reasoning of very short and very long clauses in the formula.
We consider three cases:
the first one assuming the equal share of all the lengths is to be a starting point for two subsequent cases,
the second one corresponds to a situation when rather long clauses are dominant,
and the third one when rather short clauses are dominant.

Problem \textbf{P6}
tests the impact of varying the distribution of liveness and safety clauses, 
previously assumed equal. 
While a 50:50 split reflects the natural balance in system design, 
analysis, and implementation, exploring other quantitative ratios provides further insights.

Problem \textbf{P7} arises when we need to test specific properties of a behavioural model, 
such as liveness or safety, using simple formulas related to the analysed model.
More generally, given behavioural models expressed as formulas, 
we use implication statements to check whether a required property -- 
typically a simple one -- is satisfied. 
Depending on the context, 
this verification applies to either a conjunction or a disjunction of these formulas.

Problem \textbf{P8} is slightly different. It relates to the logical square~\cite{Demey-2015}, 
a well-known concept in the literature, 
but is also applied when testing various alternative variants of behavioural models.

All eight problem sets were generated algorithmically using controlled parameters 
(clause length, atom ratio, etc.), 
enabling consistent structural variation; templates and scripts are available upon request.

\section{Experimental Results and Evaluation}
\label{sec:results-evaluation}

We evaluated the performance of First-Order Logic (FOL) theorem provers and 
compared their efficiency with a Propositional Linear Temporal Logic (PLTL) prover. 
Based on a systematic review of available solvers, 
we selected \emph{Prover9}~\cite{Prover9-tool} and 
\emph{SPASS/MSPASS}~\cite{Weidenbach-etal-2009} for FOL, 
while \emph{InKreSAT}~\cite{Kaminski-Tebbi-2013} was chosen for PLTL. 
The three provers operate with distinct input formats: 
SPASS requires TPTP, Prover9 uses LADR and InKreSAT utilises InTohyLo.
We executed all tests~\cite{Semczyszyn-2021} on a PC with 
an Intel Core i5-6400 (2.70 GHz), 
16GB RAM DDR4, running Ubuntu 18.04.4 LTS (Bionic Beaver). 
Inference computations were interrupted if they exceeded the 300-second timeout.

\begin{figure}[!htb]
\centering
\includegraphics[width=.8\columnwidth]{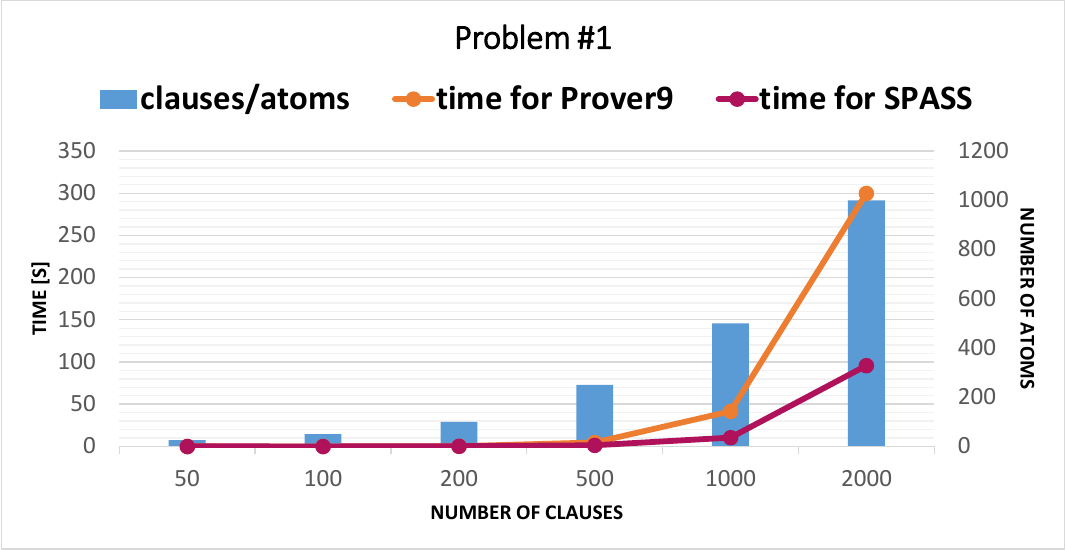}\\
\includegraphics[width=.8\columnwidth]{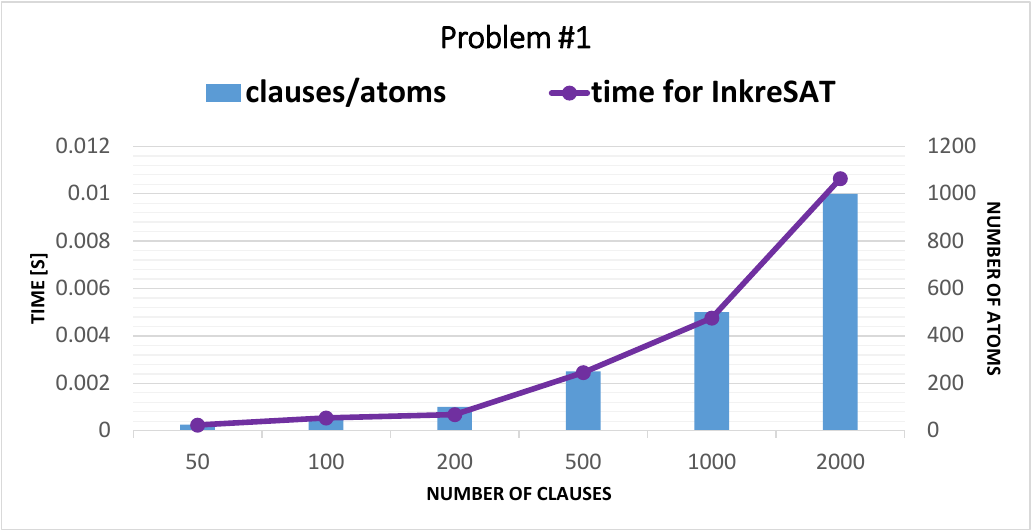}
\caption{Problem \#1, clauses against time for FOL provers (top) and InKreSAT (bottom)}
\label{fig:problem1}
\end{figure}

\begin{figure}[!htb]
\centering
\includegraphics[width=.8\columnwidth]{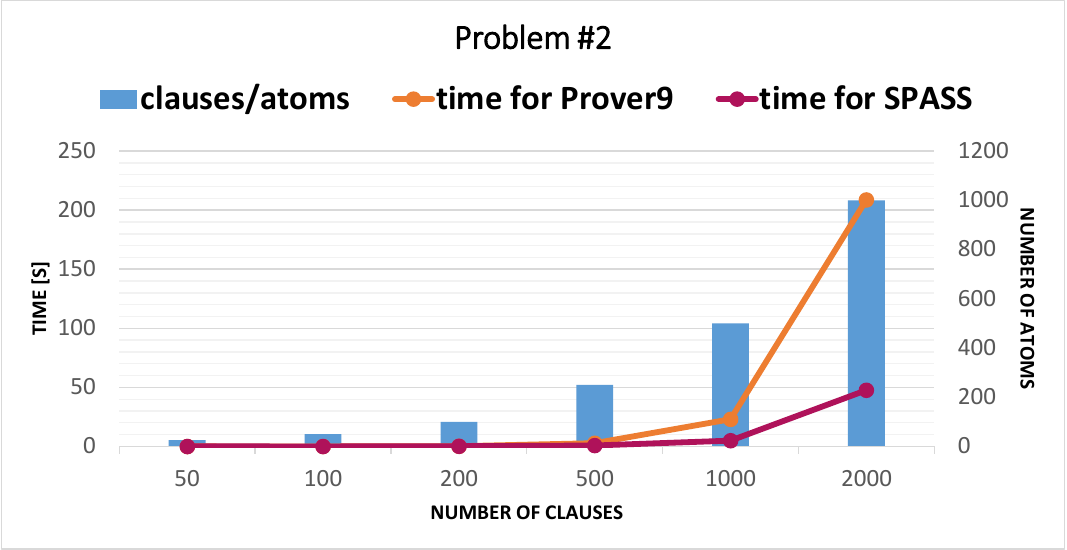}\\
\includegraphics[width=.8\columnwidth]{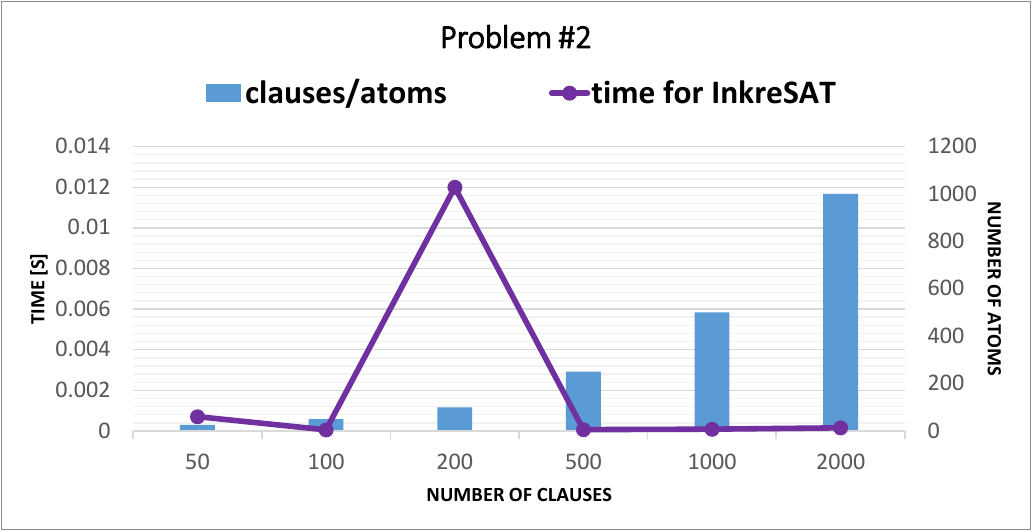}
\caption{Problem \#2, clauses against time for FOL provers (top) and InKreSAT (bottom)}
\label{fig:problem2}
\end{figure}

Figure~\ref{fig:problem1} 
presents the results of Problem \#1 
testing in terms of computational time. 
Prover9 timed out on larger formulas (2000 clauses), 
whereas SPASS and InKreSAT completed within limits, highlighting scalability differences.

\begin{figure}[!htb]
\centering
\includegraphics[width=.8\columnwidth]{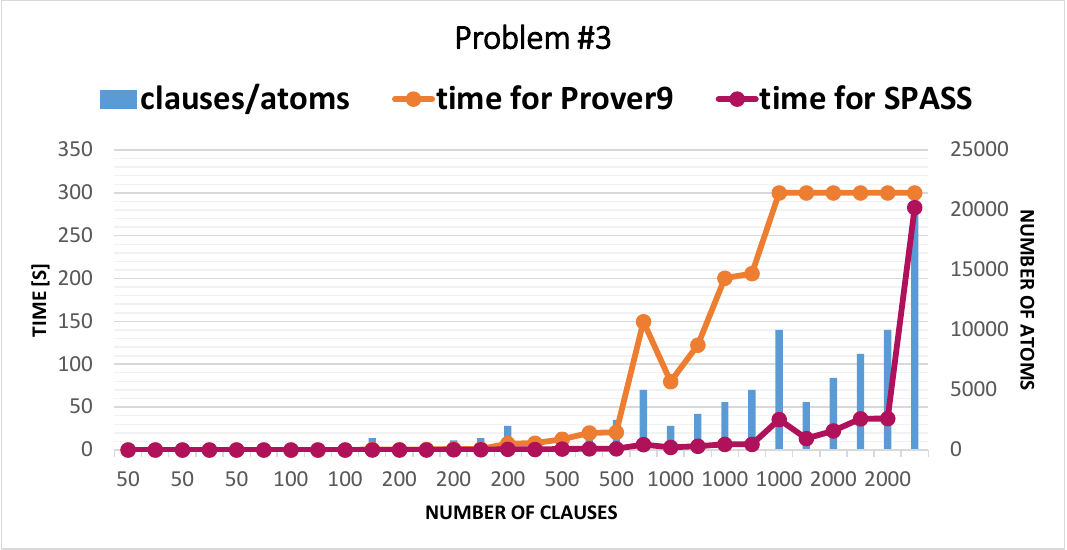}\\
\includegraphics[width=.8\columnwidth]{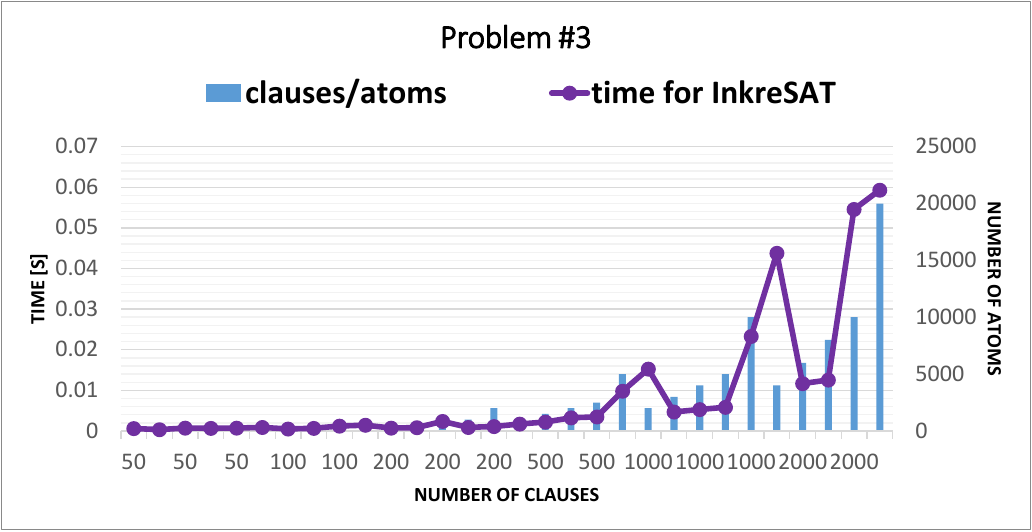}\\
\includegraphics[width=.8\columnwidth]{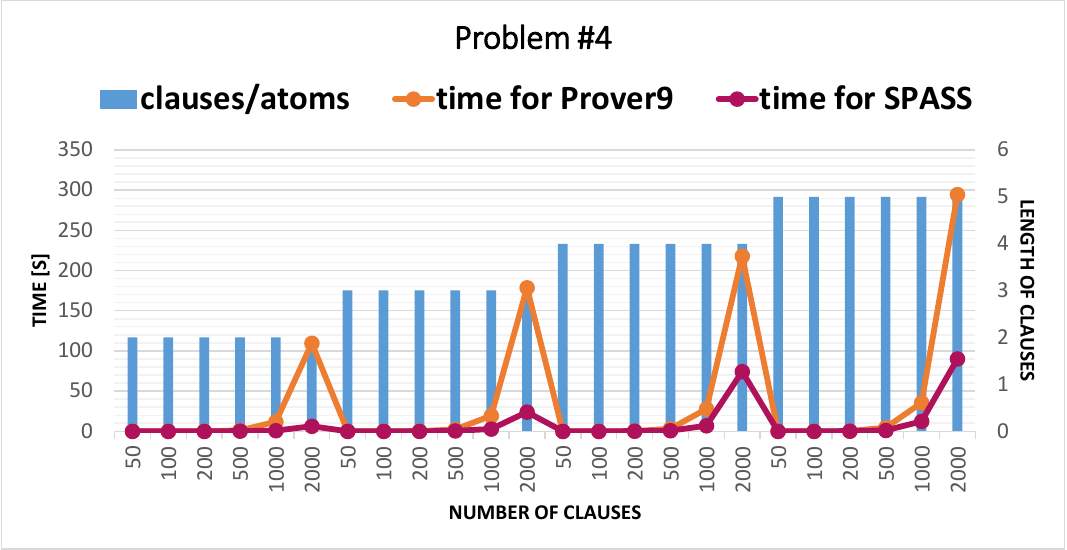}\\
\includegraphics[width=.8\columnwidth]{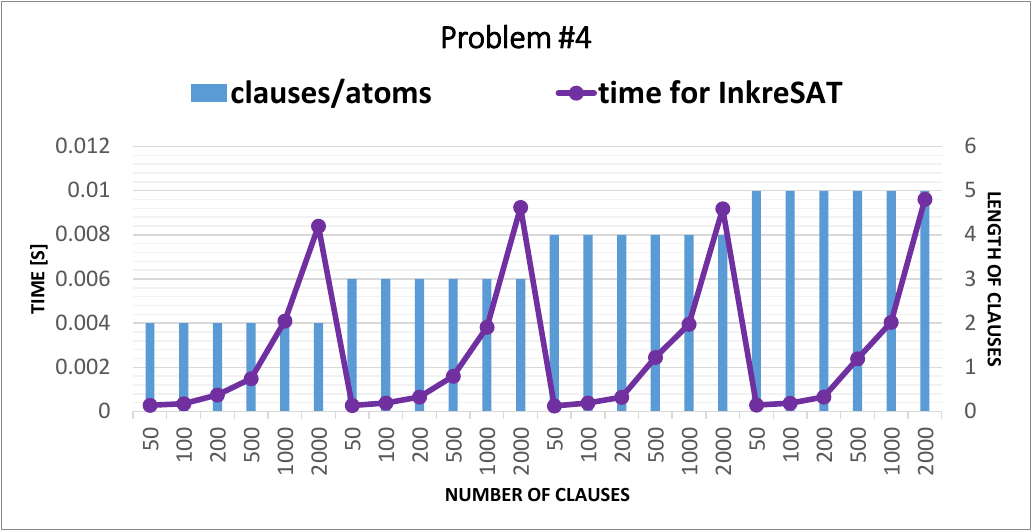}
\caption{Problem \#3 (top) and \#4 (bottom), clauses against time for FOL provers and InKreSAT}
\label{fig:problem3-4}
\end{figure}

Figure~\ref{fig:problem2} presents the results of Problem \#2 testing, 
revealing substantial differences compared to Problem \#1. 
These differences stem from a distinct approach to formula generation. 
Unlike in the previous case, 
clause quantities for specific length groups are not predefined; 
instead, clause lengths follow a Poisson distribution, 
fluctuating around an expected value. 
This distribution aligns with typical specification scenarios, 
where clauses of the approximate length 3 or 4 are the most frequently encoded, 
while those of other lengths occur less frequently.
The observed computation times exhibit similar trends to those in Problem \#1, 
though overall values are lower. 
Notably, an anomalous increase in processing time was recorded for InKreSAT 
when handling formulas with 200 clauses.

Figure~\ref{fig:problem3-4} presents the time results for testing Problems \#3 and \#4. 
The findings for Problem \#3 indicate that increasing the set of atoms 
within a formula of fixed length significantly impacts testing time 
while also prolonging the resolution process for the encoded behavioural model. 
This arises from the reduced frequency of atom occurrences within 
the formula, some may not appear at all, 
resulting in fewer contradictions, which, paradoxically, 
accelerates overall problem-solving.
Furthermore, 
formula length exerts a greater influence on testing time than 
the number of atoms involved. 
Notably, 
anomalies occur: 
for formulas with an extensive number of clauses but relatively few atoms, 
InKreSAT exhibits unexpectedly high computation times. 
Despite this atypical behaviour, InKreSAT remains substantially faster than FOL provers.

The results for Problem \#4 indicate that, 
for FOL provers, 
increasing clause length,
and thus formula length,
significantly extends testing time. 
In contrast, for InKreSAT, 
clause length has a minimal impact, 
with the number of clauses being far more influential.
All provers successfully solved the formulas, 
yet InKreSAT once again displayed notable behaviour.

\begin{figure}[!htb]
\centering
\includegraphics[width=.8\columnwidth]{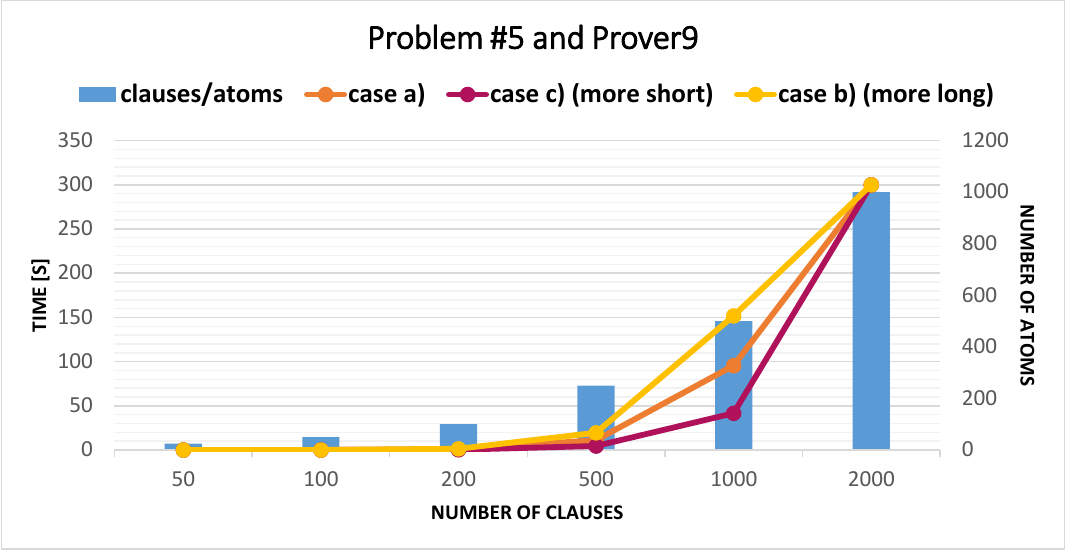}\\
\includegraphics[width=.8\columnwidth]{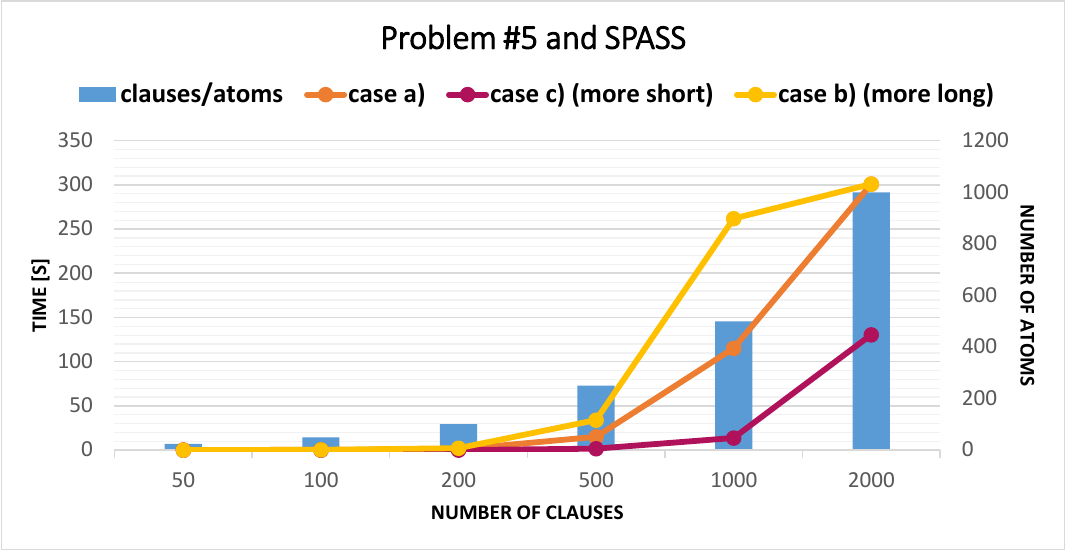}\\
\includegraphics[width=.8\columnwidth]{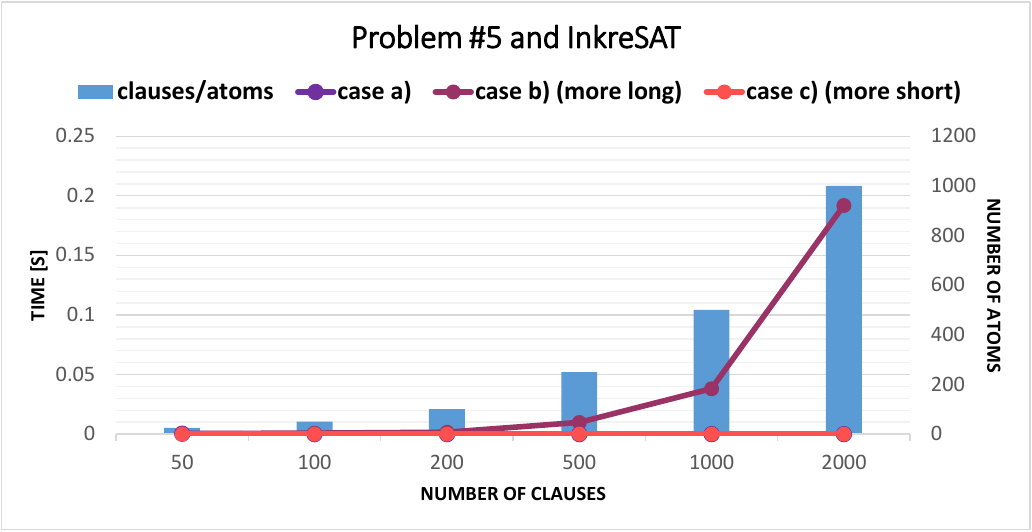}
\caption{Problem \#5 for Prover9, SPASS and InKreSAT, clauses against time}
\label{fig:problem5-prover9-spass-inkresat}
\end{figure}

Figure~\ref{fig:problem5-prover9-spass-inkresat} presents the time results for Problem \#5, 
analysed separately for each prover across three cases.
Prover9 is less impacted by average clause length than SPASS; 
they both perform better with shorter clauses. However, SPASS outperforms Prover9 on shorter formulas, 
while the reverse holds for longer ones. 
InKreSAT is significantly faster than both of them and appears proportionally affected by 
longer formulas similarly to Prover9, yet remains unaffected by medium-length formulas.

\begin{figure}[!htb]
\centering
\includegraphics[width=.8\columnwidth]{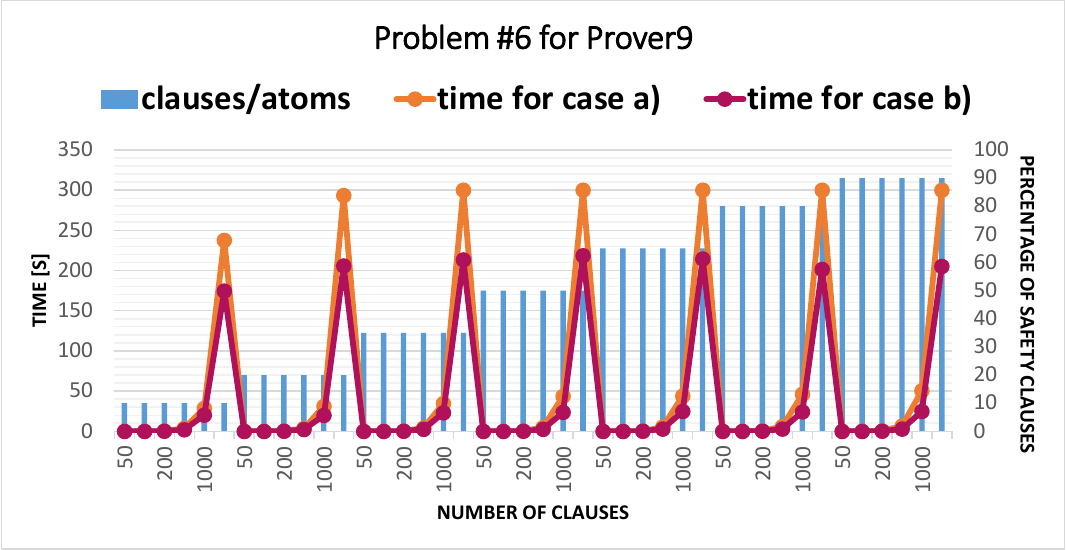}\\
\includegraphics[width=.8\columnwidth]{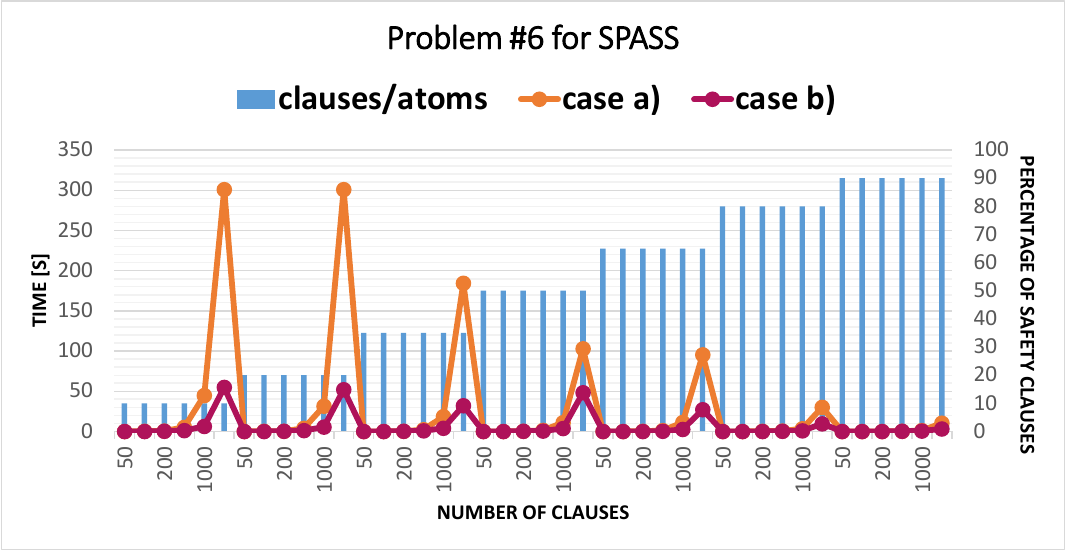}\\
\includegraphics[width=.8\columnwidth]{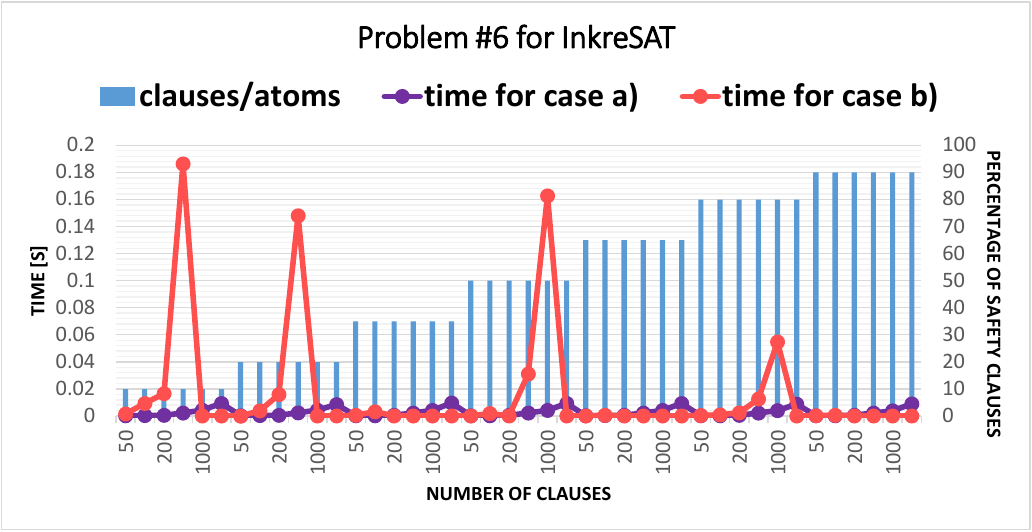}
\caption{Problem \#6 for FOL provers and InKreSAT, clauses against time}
\label{fig:problem6}
\end{figure}

Figure~\ref{fig:problem6} presents the time results for Problem \#6, 
examining the impact of different distributions of liveness and safety clauses, 
as well as two approaches to clause length.
For Poisson distribution, both FOL provers achieve better performance. 
As in previous tests, InKreSAT generally performs better with Poisson distribution but 
occasionally exhibits significantly worse results compared to even distribution. 
Prover9 performs best with a majority of liveness clauses, 
while SPASS excels with a majority of safety clauses. 
However, under Poisson distribution, SPASS maintains superior performance regardless of clause type distribution. 
In contrast, InKreSAT shows no clear correlation with the proportion of liveness and safety clauses.

\begin{figure}[!htb]
\centering
\includegraphics[width=.8\columnwidth]{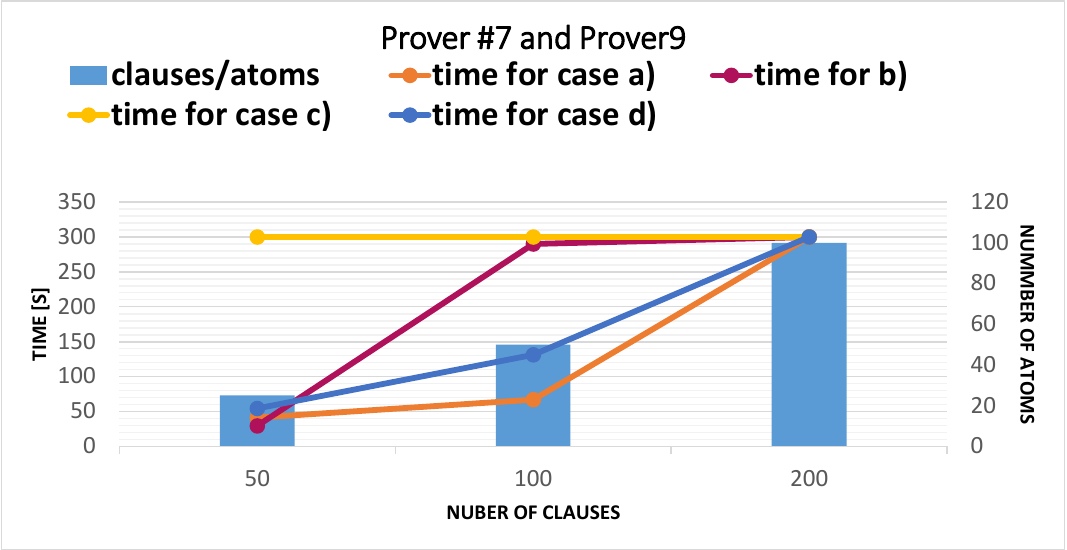}\\
\includegraphics[width=.8\columnwidth]{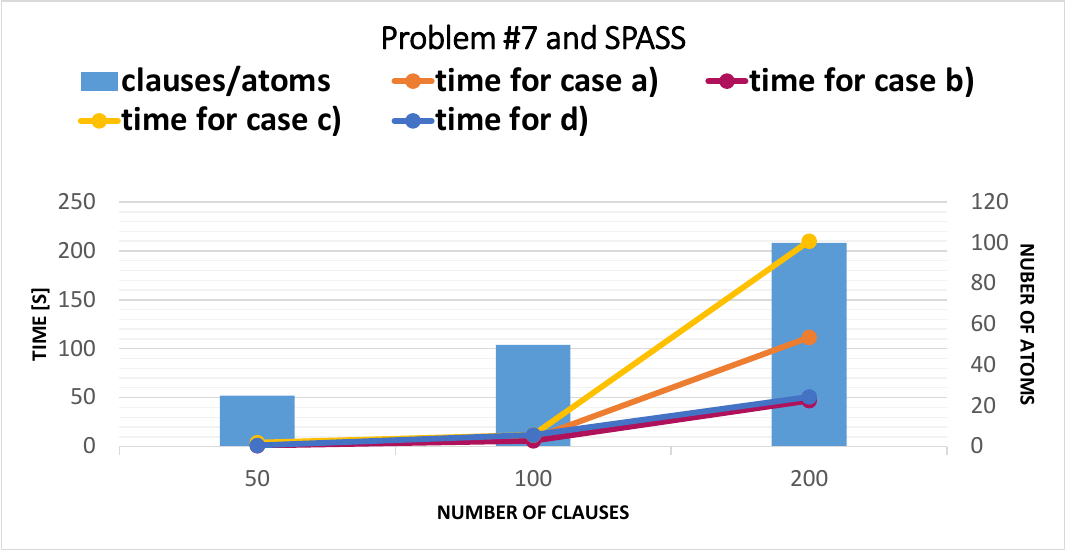}
\caption{Problem \#7 for both provers, clauses against time}
\label{fig:problem7}
\end{figure}

Figures~\ref{fig:problem7} present the time results for Problem \#7, 
which examines the satisfiability of certain properties across multiple models, 
connected by disjunction or conjunction. Each subsequent model is expressed as 
a 200-clause formula, potentially resulting in a tested formula with up to 600 clauses.
SPASS successfully solved all instances, 
whereas Prover9 frequently timed out, particularly when models were conjunction-connected. 
Overall, disjunction-connected formulas are processed more efficiently, 
similarly to formulas with Poisson distribution.

\begin{figure}[!htb]
\centering
\includegraphics[width=.8\columnwidth]{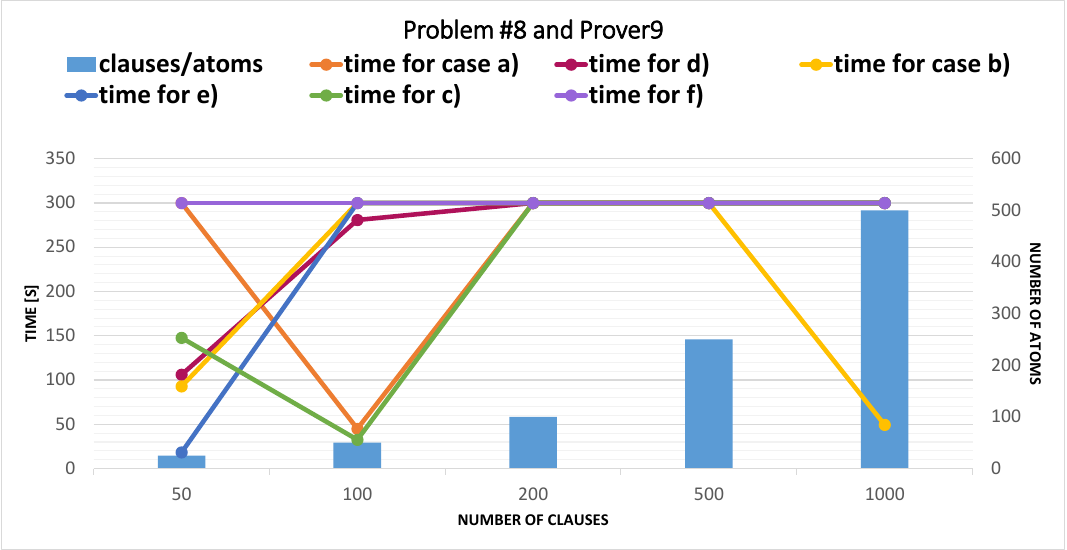}\\
\includegraphics[width=.8\columnwidth]{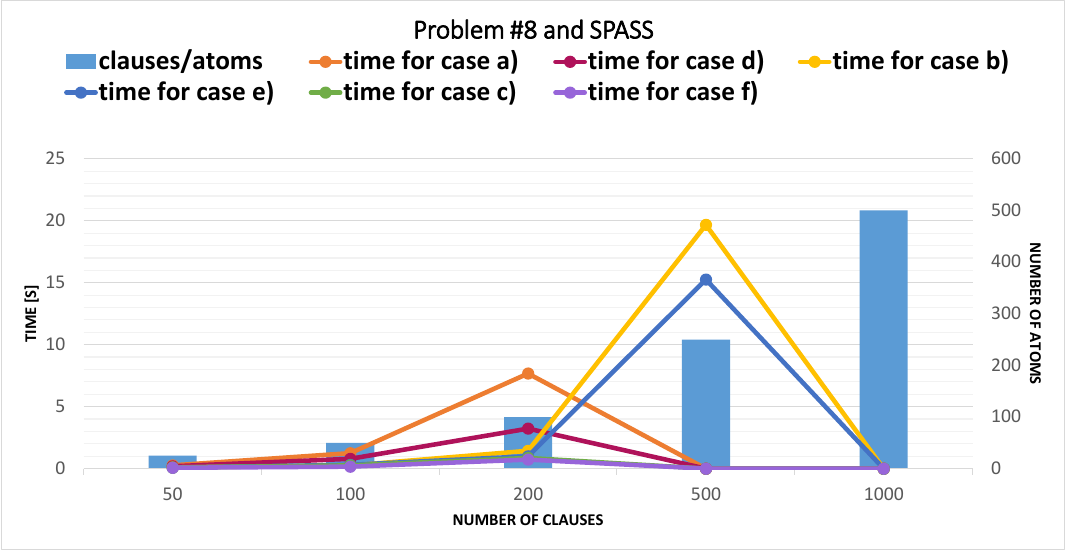}
\caption{Problem \#8 for both provers, clauses against time}
\label{fig:problem8}
\end{figure}

Figures~\ref{fig:problem8} present the time results for Problem \#8, 
comparing two models based on logical square relations. 
They also confirm the experiments~\cite{Klimek-2024-ASE-ASYDE}, 
but now with larger sets of formulas. 
Each behavioural model can be expressed with up to 1000 clauses, 
resulting in tested formulas of up to 2000 clauses.
These complex formulas could undergo preprocessing, but the study focused on direct processing as defined in sub-Problems \#8.a to \#8.c. SPASS solved all instances, occasionally exceeding memory limits, while Prover9 frequently timed out. Problem \#8.c (subalternated) had the lowest computational cost, whereas Problem \#8.a (contradictory) required significantly more time.

\begin{table}[!htb]
\caption{Time results for Problems from \#1 to \#6 for formulas with 50--500 clauses,
Prover9 \& SPASS (top) and InKreSAT (bottom).
When both extreme columns are rejected, then the average testing times are calculated.
Green marking denotes the selection of one prover for shorter formulas and another for longer formulas}
{\small
\begin{minipage}{1.1\columnwidth}
\centering
\begin{tabular}{|l|l|l|l|l|l|}
\hline
\multicolumn{6}{|c|}{\textbf{Prover9 \& SPASS}}\\
\hline
 & & \multicolumn{4}{c|}{Number of clauses in a formula}\\
\cline{3-6}
 &  & \multicolumn{1}{c}{50} & \multicolumn{1}{|c|}{\tikzmarknode{r1}{~}100} & \multicolumn{1}{c|}{200\tikzmarknode{r2}{~}} & \multicolumn{1}{c|}{500} \\
\cline{3-6}
 & & time [s] & time [s] & time [s] & time [s] \\
\hline\hline
\multirow{2}{*}{~P1~} & ~Prover9~ & 0.0167 & \cellcolor{green}0.05 & 0.0333 & 4.6767 \\
\cline{2-6}
                               & ~SPASS & 0.0533 & 0.0767 & \cellcolor{green}0.2067 & 1.5333 \\
\hline
\multirow{2}{*}{~P2~} & ~Prover9 & 0.01 &  \cellcolor{green}0.03 & 0.1467  & 2.7033 \\
\cline{2-6}
                               & ~SPASS & 0.04 & 0.0567 & \cellcolor{green}0.09 & 0.68 \\
\hline
%
\multirow{2}{*}{~P3~} & ~Prover9 & 0.0333 & \cellcolor{green}0.208 & 2.1167 & 15.255 \\
\cline{2-6}
                               & ~SPASS & 0.0547 & 0.106 & \cellcolor{green}0.328 & 2.2227 \\
\hline
\multirow{2}{*}{~P4~} & ~Prover9 & 0.01 & \cellcolor{green}0.02 & 0.0733 & 1.2333 \\
\cline{2-6}
                               & ~SPASS & 0.4 & 0.05 & \cellcolor{green}0.06 & 0.2233 \\
\hline
\multirow{2}{*}{~P5~} & ~Prover9 & 0.0267 & \cellcolor{green}0.1211 & 0.8511 & 11.7944 \\
\cline{2-6}
                               & ~SPASS & 0.0767 & 0.2711 & \cellcolor{green}1.0689 & 16.9989 \\
\hline
\multirow{2}{*}{~P6~} & ~Prover9 & 0.0167 & \cellcolor{green}0.0433 & 0.2867 & 3.5567 \\
\cline{2-6}
                               & ~SPASS & \tikzmarknode{r4}{~} 0.06 & 0.1 & \cellcolor{green}0.42 \tikzmarknode{r3}{~} & 5.1433 \\
\hline
\hline
\multicolumn{2}{|r|}{averages} & & \cellcolor{yellow}0.0787 & \cellcolor{yellow}0.3622 &\\
\hline
\end{tabular}
\begin{tikzpicture}[overlay,remember picture]
\draw[red,thick] ([yshift=6.7pt]$(r1)!-0.19!(r2)$) rectangle ([yshift=-2.5pt]$(r3)!-0.13!(r4)$);
\end{tikzpicture}
%
\begin{tabular}{|l|l|l|l|l|}
\hline
\multicolumn{5}{|c|}{\textbf{InKreSAT}}\\
\hline
 & \multicolumn{4}{c|}{Number of clauses in a formula}\\
\cline{2-5}
 &  \multicolumn{1}{c}{50} & \multicolumn{1}{|c|}{\tikzmarknode{r5}{~}100} & \multicolumn{1}{c|}{200\tikzmarknode{r6}{~}} & \multicolumn{1}{c|}{500} \\
\cline{2-5}
 & time [s] & time [s] & time [s] & time [s] \\
\hline\hline
~P1~ & 0.000238 & 0.000540 & 0.000681 & 0.002454 \\
\hline
~P2~ & 0.000699 & 0.000054 & 0.011994  & 0.000069 \\
\hline
~P3~ & 0.000601 & 0.000939 & 0.001183 & 0.004092 \\
\hline
~P4~ & 0.000271 & 0.000370 & 0.000673 & 0.001977 \\
\hline
~P5~ & 0.000328 & 0.000304 & 0.000588 & 0.003353 \\
\hline
~P6~ & \tikzmarknode{r7}{~}0.000353 & 0.001680 & 0.002865\tikzmarknode{r8}{~} & 0.028196 \\
\hline
\hline
\multicolumn{1}{|r|}{} & & \cellcolor{yellow}0.000648 & \cellcolor{yellow}0.002997 &\\
\hline
\end{tabular}
\begin{tikzpicture}[overlay,remember picture]
\draw[red,thick] ([yshift=6.7pt]$(r5)!1.23!(r6)$) rectangle ([yshift=-2.5pt]$(r7)!0.34!(r8)$);
\end{tikzpicture}
\end{minipage}
}
\label{tab:testting-times}
\end{table}

Table~\ref{tab:testting-times} 
presents a summary of testing times for the various problems and formulae under consideration. It is clearly observable that both FOL provers perform well across all formulae. 
In the case of shorter formulae, Prover9 generally outperforms SPASS.
The results obtained for InKreSAT exhibit some outliers, as previously discussed. 
Nevertheless, 
all results remain fully acceptable, with the average approximate performance which is
one hundred times superior to that of the FOL provers. 
This can be attributed, at least to a certain extent, 
to the significantly lower memory requirements of Prover9, 
which were not explicitly illustrated due to space constraints in this article.
Prover9 achieves superior execution times for shorter formulae, 
particularly those comprising 200--500 clauses. 
In contrast, SPASS demonstrates better performance for substantially longer formulae. 
Notably, 
InKreSAT consistently outperforms both FOL provers across all formulae in terms of execution time.
Although the Poisson distribution—characterised by its intrinsic irregularities in 
length distribution—leads to a substantial improvement in execution times, 
it occasionally introduces abrupt and irregular decreases in efficiency for InKreSAT. 
Nevertheless, even in such cases, InKreSAT remains significantly more efficient than the FOL provers.

Assuming that the objective is to develop 
an Integrated Development Environment (IDE)-class tool with built-in interaction based on deductive reasoning about behavioural models, the following statement can be formulated.
\begin{claim}
\label{the:provers-4-ide}
Automated reasoning on behavioural models should be completed within 1–2 seconds 
using FOL provers to support real-time feedback in IDEs. 
Our results demonstrate this is feasible for formulas up to a medium size, 
aligning with needs in interactive model validation.
For the PLTL prover, the obtained results demonstrate an improvement by a factor of 100.
\end{claim}
\begin{claim}
\label{the:provers-fol-vs-pltl}
InKreSAT consistently outperforms FOL provers, 
offering near-instant validation via efficient heuristics and incremental solving. 
This makes it a strong candidate for integration in CI/CD pipelines and 
AI-assisted development tools.
\end{claim}
Despite minor irregularities, InKreSAT outperforms other provers by an order of magnitude, proving its efficiency even in complex formulas. Its speed advantage makes it highly effective for safety-critical verification, where both performance and stability matter.

\section{Related works}


Over the years, 
various structured collections have been developed to test 
the efficiency and applicability of theorem provers and SAT solvers. 
Pelletier~\cite{Pelletier-1986} provided an early benchmark suite for first-order logic (FOL), consisting of structured logical problems widely used in theorem proving research. Sutcliffe~\cite{Sutcliffe-2017} expanded this with 
the TPTP (Thousands of Problems for Theorem Provers) library, 
a standard for evaluating provers across logic domains. 
While our approach targets logic-based behavioural modelling, 
the TPTP benchmark is oriented toward general theorem proving. 
Mitchell et al.~\cite{Mitchell-etal-1992} analysed SAT problem structures, 
identifying conditions that make some instances significantly harder. 
Their focus is benchmarking; ours targets logic-based verification in software engineering contexts.

To sum up, 
our study constructs a structured problem catalogue explicitly designed for 
evaluating automated logical specification techniques in behavioural modelling. 
This work enhances empirical validation frameworks for 
IDE-oriented and AI-driven software verification tools, 
addressing scalability and applicability concerns in practical use cases. 
Earlier prototyping efforts lacked structured IDE integration, 
limiting their utility in such 
contexts~\cite{Klimek-Szwed-2013-FedCSIS}.

\section{Conclusion and further works}
\label{sec:conclusion}

We present a logic-oriented framework, focusing on the coherent idea and technical feasibility of F-IDE. 
Nearly two thousand tests ($1890 = 210$ tasks $\times 3$ solvers $\times 3$ attempts) 
were conducted on the proposed logical problems for theorem provers.

Automated theorem proving could be embedded in CI/CD pipelines for real-time logical verification, 
and on-the-fly model validation in IDEs could provide instant feedback, reducing post-hoc efforts. 
Short-to-medium sized formulae (e.g., P1--P3) resemble behavioural contracts in 
microservice deployments, 
while larger implication-based problems (e.g., P7--P8) 
reflect cross-model verification in architecture-level design.

Our findings validate prior studies while exposing scalability limitations and solver inefficiencies. 
Clause length, atom ratio, and logical structure critically affect performance; 
fixed lengths improve predictability, while implication-based verification remains demanding. 
Future work should address dataset extension and AI-assisted heuristic optimisation. 
The dataset, generation scripts, and raw results will be made publicly available 
upon acceptance via a dedicated repository.

%
%


\bibliographystyle{ACM-Reference-Format}
\bibliography{bib-rk,bib-rk-main,bib-rk-tools}

\end{document}